\documentclass[11pt]{article}

\usepackage[preprint]{acl}

\usepackage{times}
\usepackage{makecell}
\usepackage{latexsym}
 \usepackage{amsmath}
 \usepackage{dblfloatfix}
\usepackage{amssymb}
\usepackage{booktabs}
\usepackage{multirow}
\usepackage{adjustbox}
\usepackage{longtable}
\usepackage{caption}
\usepackage{subcaption}
\usepackage{graphicx}
\usepackage{caption}
\usepackage[T1]{fontenc}

\usepackage[utf8]{inputenc}

\usepackage{microtype}

\usepackage{inconsolata}

\usepackage{graphicx}

\title{Learning to Select: Query-Aware Adaptive Dimension Selection for Dense Retrieval}

\author{Zhanyu Wu \\
  Beihang University, China \\
  \texttt{wuzy24@act.buaa.edu.cn} \\\And
  Richong Zhang \\
  Beihang University, China \\
  \texttt{zhangrc@act.buaa.edu.cn} \\\And
  Zhijie Nie \\
  Beihang University, China \\
  \texttt{niezj@act.buaa.edu.cn} }

\begin{document}
\maketitle
\begin{abstract}
Dense retrieval represents queries and documents as high-dimensional embeddings, but these representations can be redundant at the query level: for a given information need, only a subset of dimensions is consistently helpful for ranking. Prior work addresses this via pseudo-relevance feedback (PRF) based dimension importance estimation, which can produce query-aware masks without labeled data but often relies on noisy pseudo signals and heuristic test-time procedures. In contrast, supervised adapter methods leverage relevance labels to improve embedding quality, yet they learn global transformations shared across queries and do not explicitly model query-aware dimension importance. We propose a Query-Aware Adaptive Dimension Selection framework that \emph{learns} to predict per-dimension importance directly from query embedding. We first construct oracle dimension importance distributions over embedding dimensions using supervised relevance labels, and then train a predictor to map a query embedding to these label-distilled importance scores. At inference, the predictor selects a query-aware subset of dimensions for similarity computation based solely on the query embedding, without pseudo-relevance feedback. Experiments across multiple dense retrievers and benchmarks show that our learned dimension selector improves retrieval effectiveness over the full-dimensional baseline as well as PRF-based masking and supervised adapter baselines.
\end{abstract}

\section{Introduction}

Dense retrieval has become central to modern IR, mapping queries and documents into high-dimensional vector spaces for contextual matching \cite{karpukhin2020dense,xiong2020approximate}. Compared to classic sparse approaches like BM25 \cite{robertson2009probabilistic} and learned sparse models \cite{formal2021splade,formal2021spladev2}, dense embeddings often yield substantial gains by capturing higher-level semantic similarity. However, these high-dimensional representations can be redundant at the \emph{query} level: for a given information need, only a subset of dimensions contributes meaningfully to relevance, while others may be neutral or even harmful. This motivates selecting informative dimensions per query to improve effectiveness.

Recent work tackles this dimension redundancy issue directly via \emph{Dimension Importance Estimation} \cite{faggioli2024dimension,faggioli2025getting,campagnano2025unveiling}. DIME-style approaches estimates per-query dimension importance from pseudo-relevance feedback \cite{rocchio1971relevance} or LLM-generated pseudo-documents by scoring dimensions according to their alignment with pseudo-positives, sometimes also using pseudo-negatives in a contrastive manner \cite{d2025eclipse}. While these methods yield query-aware masks, they rely on pseudo-labels that can be noisy, and they are implemented as separate heuristic stages at inference time rather than being trained to directly exploit available supervised relevance labels.

A complementary line of work instead uses \emph{learned adapters} to directly improve representations with supervised signals \cite{yoon2024search}.
Such adapters are trained on labeled retrieval data to reshape the embedding space and often achieve substantially better full-dimensional performance than the original encoder. However, these adapters learn a \emph{global} transformation shared by all queries and documents, so supervised signals are used to model corpus-level structure rather than query-aware patterns.

\begin{figure}[t]
  \centering
  \includegraphics[width=\linewidth]{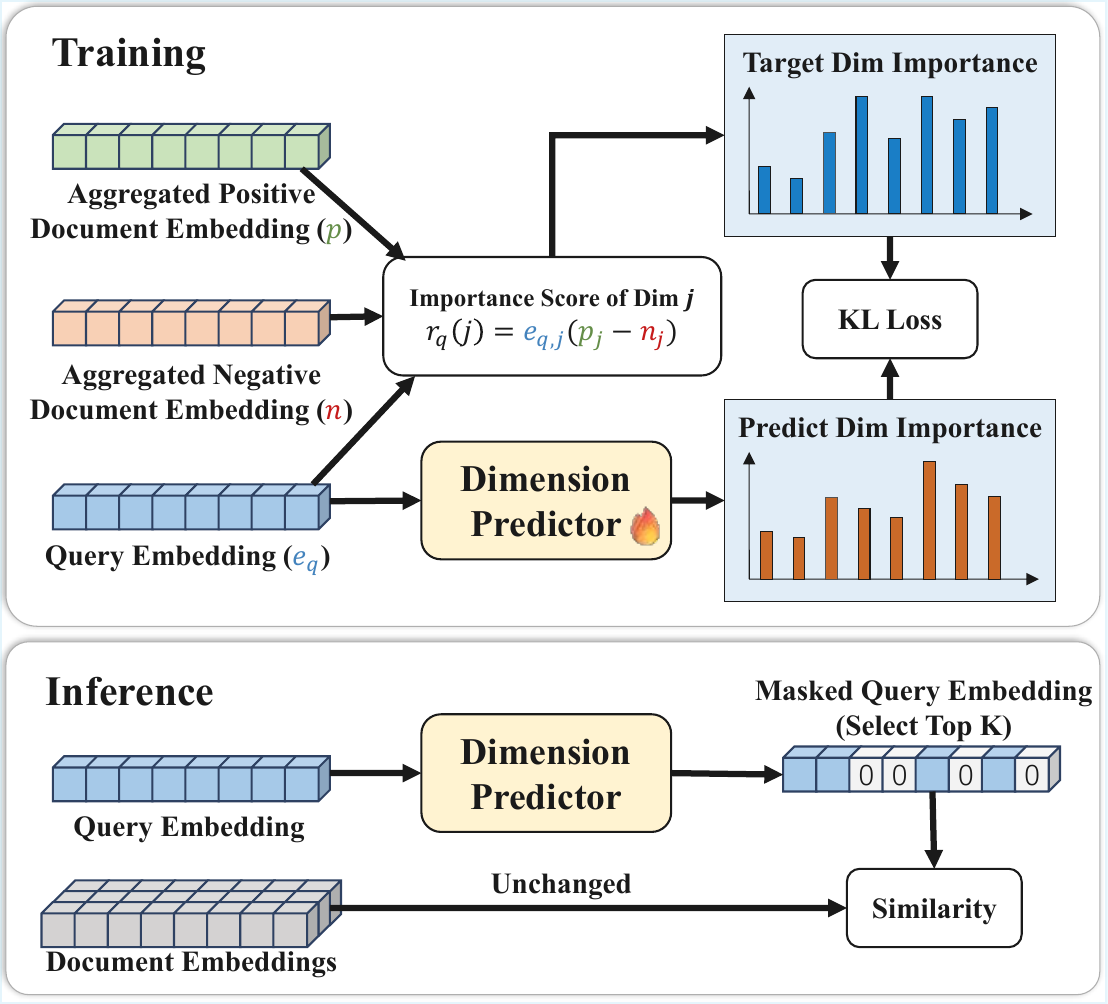}
  \caption{
Our query-aware dimension selection pipeline. We construct a per-query dimension-importance target by contrasting aggregated embeddings of relevant documents against aggregated embeddings of hard negatives. A query-only predictor is trained to match this target distribution via KL divergence. At inference, the predicted importance selects top‑k query dimensions (masking the rest), while document embeddings and the ANN index remain unchanged.
  }
  \label{fig:method_overview}
\end{figure}

Motivated by these limitations, we aim to learn a supervised predictor of query-aware dimension importance, avoiding noisy pseudo-labeling while retaining the benefits of learned supervision. We propose a query-aware dimension selection framework that learns to predict dimension importance from supervised retrieval signals. The core idea is to distill \emph{oracle} importance scores for each query from relevance labels, and then train a predictor—a small fully connected module on top of frozen embeddings—to approximate these scores from query semantics alone. At inference, the predictor outputs per-dimension importance over the query embedding; we then mask low-importance dimensions in the query vector and compute similarity with unmodified document embeddings. Figure~\ref{fig:method_overview} summarizes our query-aware dimension selection pipeline. 

Our approach improves retrieval effectiveness by learning query-aware dimension importance directly from supervised relevance signals, replacing noisy pseudo-feedback and heuristic test-time estimation with offline label-distilled learning. This enables fine-grained, label-informed selection patterns that pseudo-feedback methods and global adapter transformations can miss. 
For deployment, we apply the learned mask by zeroing out unselected query dimensions, keeping each query a fixed $D$-dimensional vector and leaving document embeddings and ANN indexes (e.g., FAISS~\cite{johnson2019billion}) unchanged.

In summary, our main contributions are:
\begin{itemize}
\item To the best of our knowledge, we are the first to learn per-query, dimension-level importance from supervised relevance labels and use it for query-only masking at inference.
\item We learn dimension importance directly from supervised relevance labels by distilling per-query oracle importance targets and training the predictor offline to match them, avoiding pseudo-feedback estimation.
\item We show that query-side top-$k$ dimension masking yields consistent retrieval improvements across multiple benchmarks and dense retrievers, while keeping document embeddings and ANN indexes unchanged.
\end{itemize}
\section{Related Work}
\paragraph{Dense retrieval}
Dense retrieval employs neural encoders to transform
queries and documents into fixed-length vector representations, computing their semantic relevance through functions like cosine similarity.  Let $q$ denote a query and $d$ denote a document. A dual-encoder produces embeddings $e_q \in \mathbb{R}^D$ and $e_d \in \mathbb{R}^D$:
\[
e_q = f(q), \quad e_d = g(d),
\]
where $f$ and $g$ are pre-trained encoders. 

Relevance is estimated via a similarity function $s(e_q, e_d)$, commonly cosine similarity:
\[
s(e_q, e_d) = \frac{e_q^\top e_d}{\|e_q\|_2 \, \|e_d\|_2}
,
\]
where $||e||_2$ is the Euclidean norm.

In our setting, we pre-normalize all embeddings with $\ell_2$ normalization.
Specifically, we replace $e$ with $\hat{e} = \frac{e}{\|e\|_2}$ for both queries and documents.
 
\paragraph{Search Adapter}
Search adapter \cite{yoon2024search} learns a post-embedding transform on top of a frozen encoder using supervised relevance signals.
Given query and document embeddings $e_q, e_d \in \mathbb{R}^D$, the adapter defines
\[
\tilde{e}_q \;=\; A e_q, \qquad \tilde{e}_d \;=\; A e_d,
\]
where $A \in \mathbb{R}^{D \times D}$ is a trainable projection, and retrieval is performed using the adapted embeddings $\tilde{e}_q, \tilde{e}_d$. While search adapter uses a trainable linear projection to transform embeddings for retrieval, our linear layer is instead used to select the most informative dimensions.

\paragraph{Matryoshka Embedding and Matryoshka Adapter}
Matryoshka representation learning (MRL) \cite{kusupati2022matryoshka} trains the encoder so that embeddings are \emph{nested} across dimensions, enabling prefix truncation while preserving ranking quality. 
Building on this idea, Matryoshka adapters \cite{yoon2024matryoshka,zhang2025smec} learn $A$ with an MRL-style objective over a set of prefix sizes $\{D_1 < D_2 < \cdots < D_M = D\}$, encouraging the prefixes $\tilde{e}_q^{[1:D_i]}$ and $\tilde{e}_d^{[1:D_i]}$ to remain informative for retrieval. In contrast to Search Adapter, Matryoshka-style methods primarily target efficiency by ensuring strong retrieval quality under low-dimensional prefix truncation, while their full-dimensional peak performance is typically comparable to Search Adapter. 
Unlike Matryoshka \cite{kusupati2022matryoshka} methods that trade off accuracy for efficiency, our primary goal is effectiveness. We employ dimension selection as a denoising mechanism to remove harmful redundancy.

\paragraph{Dimension Importance Estimation (DIME).}
DIME \cite{faggioli2024dimension,faggioli2025getting,campagnano2025unveiling} adaptively selects, for each query, a subset of important dimensions to compute similarity only on those coordinates, improving retrieval effectiveness. Let the query and document embeddings be $e_q, e_d \in \mathbb{R}^D$.

For each dimension $j$, DIME estimate importance from query–positive alignment:
\[
\pi_q(j) \;\propto\; e_{q,j}\,\bigg(\frac{1}{|D^+(q)|}\sum_{d\in D^+(q)} e_{d,j}\bigg),
\]
where $D^+(q)$ is a set of relevant documents for $q$. DIME obtains $D^+(q)$ via (i) LLM-generated pseudo-relevant documents, or (ii) pseudo-relevance feedback (PRF) using the top-$m$ retrieved documents, treating them as positives. Recent work further extends this paradigm by also exploiting pseudo-irrelevant information: Eclipse \cite{d2025eclipse} introduces contrastive dimension importance estimation that leverages both pseudo-positives and pseudo-negatives, encouraging dimensions that better discriminate relevant from non-relevant feedback documents.

For a target dimension $k$, let
\[
S_k(q) \;=\; \mathrm{Top}\text{-}k\big(\{\pi_q(j)\}_{j=1}^D\big),
\]
and compute similarity restricted to $S_k(q)$. Empirically, such dimension filtering improves ranking metrics without modifying the underlying index.

\section{Method}

We learn a per-query dimension-importance predictor that identifies which coordinates of the query embedding are most relevant for matching. The procedure has two steps: calculate oracle dimension importance and train the dimension importance predictor.

\subsection{Calculate Oracle Dimension Importance}
Given training triples $(q,d,y)$ where $y$ denotes the multi-level supervised relevance label of document $d$ to query $q$,  with embeddings $e_q, e_d \in \mathbb{R}^D$, we estimate per-query dimension importance by selecting coordinates that best discriminate relevant documents from hard negatives. Intuitively, a dimension is valuable if it contributes strongly to the query–positive similarity but weakly (or negatively) to the query–negative similarity.

For each query $q$, collect relevant documents $D^+(q)=\{d:\, y(d)>0\}$ (queries without any relevant documents are skipped) and define gain scores $g_d \;=\; 2^{y(d)} - 1$. 
We then normalize them into weights
\[
w_d \;=\; \frac{g_d}{\sum_{d' \in D^+(q)} g_{d'}}.
\]
The weighted positive embeddings centroid is
\[
p \;=\; \sum_{d \in D^+(q)} w_d \, e_d.
\]

Rank all documents by similarity $s(d)=e_d \cdot e_q$, take the top-$K$ non-relevant items, and uniformly sample $M$ negatives to form $D^-(q)$ (with $K$ and $M$ as tunable hyperparameters). The mean negative embedding is
\[
n \;=\; \frac{1}{|D^-(q)|}\sum_{d \in D^-(q)} e_d.
\]

For each dimension $j \in \{1,\dots,D\}$, compute a raw discrimination score as the positive–negative alignment difference on that coordinate,
\[
r_q(j) \;=\; e_{q,j}\, (p_j -\,n_j),
\]
which quantifies how strongly dimension $j$ supports relevant matches while repelling negatives. We then apply a temperature-scaled softmax over dimensions to obtain a dimensions importance distribution,
\[
\pi_q \;=\; \mathrm{softmax}\!\big(r_q/\tau\big),
\]
where $\tau>0$ controls sharpness. 

\subsection{Train Dimension Importance Predictor}
We use a fully connected layer $f_\theta: \mathbb{R}^D \to \mathbb{R}^D$. The predictor takes the frozen query embedding $e_q$ as input, and outputs logits $\ell=f_\theta(e_q)$ and log-probabilities $\log \hat{\pi}_q=\log\mathrm{softmax}(\ell)$. The training objective minimizes the KL divergence between the target distribution and the prediction:
\[
\mathcal{L}(q) \;=\; \mathrm{KL}\!\big(\pi_q \,\|\, \hat{\pi}_q\big).
\]
At inference, we select the query-aware top-$k$ dimensions according to the predicted dimensions importance $\hat{\pi}_q$.
Let
$$
\mathcal{J}_q^{(k)}=\mathrm{Top}\text{-}k(\hat{\pi}_q),\qquad
m_{q,j}^{(k)}=\mathbf{1}[j\in \mathcal{J}_q^{(k)}],
$$
and define the masked query embedding
$$
e_q^{(k)} = e_q \odot m_q^{(k)}.
$$
We then compute similarity using the masked query embedding and original document embedding:
$$
s_k(e_q,e_d)=(e_q^{(k)})^\top e_d.
$$
This is equivalent to evaluating similarity only on the selected dimensions, while keeping document embeddings and the ANN index unchanged (no retraining or reindexing).
\begin{table*}[t]
\centering
\small
\caption{
Peak retrieval effectiveness (NDCG@10; higher is better) and corresponding fraction of dimensions used (in parentheses) across all models. Superscripts indicate how each method is trained: $^{\text{u}}$ = fully unsupervised, $^{\text{p}}$ = PRF-based, $^{\text{s}}$ = supervised with relevance labels. The bottom row for each dataset block reports \textbf{Ours} at a fixed retained dimensions ratio of 30\%.
}
\label{tab:main_all_models}
\begin{adjustbox}{max width=\textwidth}
\begin{tabular}{lllllllll}
\toprule
Dataset & Method & Qwen-0.6B & Qwen-4B & Qwen-8B & OpenAI & L2V-Mistral & L2V-LLaMA & GritLM \\
\midrule
\multirow{8}{*}{SciFact}
& Baseline$^{u}$ & 0.702 (100\%) & 0.785 (100\%) & 0.783 (100\%) & 0.776 (100\%) & 0.773 (100\%) & 0.787 (100\%) & 0.786 (100\%) \\
& Cutoff$^{u}$ & 0.702 (100\%) & 0.788 (86\%) & 0.784 (98\%) & 0.776 (98\%) & 0.773 (100\%) & 0.788 (84\%) & 0.792 (60\%) \\
& Norm$^{u}$ & 0.706 (64\%) & 0.789 (40\%) & 0.785 (32\%) & 0.778 (80\%) & 0.774 (62\%) & 0.787 (92\%) & 0.791 (28\%) \\
& DIME$^{p}$ & 0.705 (98\%) & 0.787 (94\%) & 0.785 (94\%) & 0.781 (82\%) & 0.773 (100\%) & 0.788 (98\%) & 0.787 (98\%) \\
& Eclipse$^{p}$ & 0.704 (98\%) & 0.787 (98\%) & 0.785 (94\%) & 0.783 (94\%) & 0.773 (100\%) & 0.788 (98\%) & 0.787 (98\%) \\
& Adapter$^{s}$ & 0.774 (100\%) & 0.849 (100\%) & \textbf{0.883} (100\%) & 0.871 (100\%) & \textbf{0.843} (100\%) & 0.882 (100\%) & 0.883 (100\%) \\
& \textbf{Ours}$^{s}$ & \textbf{0.845} (32\%) & \textbf{0.899} (56\%) & 0.883 (32\%) & \textbf{0.897} (32\%) & 0.830 (20\%) & \textbf{0.884} (10\%) & \textbf{0.906} (40\%) \\
\cmidrule(lr){2-9}
& \textbf{Ours}@30\%$^{s}$ & 0.839 (30\%) & 0.895 (30\%) & 0.881 (30\%) & 0.897 (30\%) & 0.823 (30\%) & 0.881 (30\%) & 0.902 (30\%) \\
\midrule
\multirow{8}{*}{MS MARCO}
& Baseline$^{u}$ & 0.562 (100\%) & 0.683 (100\%) & 0.646 (100\%) & 0.681 (100\%) & 0.613 (100\%) & 0.562 (100\%) & 0.536 (100\%) \\
& Cutoff$^{u}$ & 0.562 (100\%) & 0.683 (100\%) & 0.651 (52\%) & 0.690 (74\%) & 0.623 (76\%) & 0.565 (94\%) & 0.554 (54\%) \\
& Norm$^{u}$ & 0.575 (52\%) & 0.687 (42\%) & 0.653 (62\%) & 0.688 (10\%) & 0.617 (40\%) & 0.569 (48\%) & 0.537 (86\%) \\
& DIME$^{p}$ & 0.595 (84\%) & 0.701 (14\%) & 0.664 (40\%) & \textbf{0.702} (10\%) & 0.638 (70\%) & 0.582 (64\%) & 0.566 (64\%) \\
& Eclipse$^{p}$ & 0.602 (18\%) & \textbf{0.702} (28\%) & 0.668 (84\%) & 0.700 (32\%) & \textbf{0.641} (86\%) & 0.588 (50\%) & 0.565 (76\%) \\
& Adapter$^{s}$ & 0.607 (100\%) & 0.682 (100\%) & 0.698 (100\%) & 0.680 (100\%) & 0.621 (100\%) & 0.576 (100\%) & 0.593 (100\%) \\
& \textbf{Ours}$^{s}$ & \textbf{0.609} (44\%) & 0.699 (20\%) & \textbf{0.714} (20\%) & 0.700 (34\%) & 0.638 (24\%) & \textbf{0.600} (26\%) & \textbf{0.632} (40\%) \\
\cmidrule(lr){2-9}
& \textbf{Ours}@30\%$^{s}$ & 0.601 (30\%) & 0.693 (30\%) & 0.709 (30\%) & 0.696 (30\%) & 0.634 (30\%) & 0.593 (30\%) & 0.631 (30\%) \\
\midrule
\multirow{8}{*}{NFCorpus}
& Baseline$^{u}$ & 0.377 (100\%) & 0.426 (100\%) & 0.432 (100\%) & 0.440 (100\%) & 0.407 (100\%) & 0.432 (100\%) & 0.429 (100\%) \\
& Cutoff$^{u}$ & 0.379 (86\%) & 0.426 (100\%) & 0.435 (32\%) & 0.440 (100\%) & 0.411 (62\%) & 0.434 (56\%) & 0.432 (62\%) \\
& Norm$^{u}$ & 0.379 (62\%) & 0.426 (96\%) & 0.433 (80\%) & 0.440 (90\%) & 0.410 (58\%) & 0.436 (60\%) & 0.430 (92\%) \\
& DIME$^{p}$ & 0.384 (94\%) & 0.432 (80\%) & 0.436 (86\%) & 0.449 (84\%) & 0.414 (50\%) & 0.433 (98\%) & 0.433 (24\%) \\
& Eclipse$^{p}$ & 0.384 (90\%) & 0.432 (50\%) & 0.435 (40\%) & 0.449 (78\%) & 0.415 (56\%) & 0.434 (94\%) & 0.434 (94\%) \\
& Adapter$^{s}$ & 0.371 (100\%) & 0.427 (100\%) & 0.419 (100\%) & 0.428 (100\%) & \textbf{0.417} (100\%) & 0.442 (100\%) & 0.434 (100\%) \\
& \textbf{Ours}$^{s}$ & \textbf{0.396} (62\%) & \textbf{0.451} (22\%) & \textbf{0.455} (38\%) & \textbf{0.459} (38\%) & 0.412 (54\%) & \textbf{0.445} (16\%) & \textbf{0.462} (58\%) \\
\cmidrule(lr){2-9}
& \textbf{Ours}@30\%$^{s}$ & 0.388 (30\%) & 0.447 (30\%) & 0.449 (30\%) & 0.455 (30\%) & 0.405 (30\%) & 0.442 (30\%) & 0.461 (30\%) \\
\bottomrule
\end{tabular}
\end{adjustbox}
\end{table*}

\section{Experimental Setup}

\paragraph{Overview}
We evaluate a query-aware dimension-importance predictor across multiple embedding models and datasets. For each model–dataset pair, we (i) generate L2-normalized query and document embeddings, (ii) train a predictor solely on embeddings and relevance labels, and (iii) at test time, select the top-$k$ dimensions per query and compute similarity restricted to those coordinates. Performance is reported as a function of $k$.

\paragraph{Models}
We evaluate our method on a diverse set of dense retrievers, including both encoders trained with explicit multi-scale (MRL-style) objectives and standard encoders without such constraints.
We use Qwen-Embedding models trained with the Matryoshka objective at different scales 
(Qwen-Embedding-0.6B, $d\!=\!1024$; Qwen-Embedding-4B, $d\!=\!2560$; Qwen-Embedding-8B, $d\!=\!4096$)~\cite{zhang2025qwen3}, 
and \texttt{text-embedding-3-large} from OpenAI ($d\!=\!3072$)~\cite{openai2024embedding}. 
These encoders are designed to support prefix-based truncation.
We additionally include dense encoders that are not trained with explicit multi-scale objectives: 
LLM2Vec-Mistral-7B ($d\!=\!4096$) ~\cite{behnamghader2024llm2vec}, 
LLM2Vec-Llama3-8B ($d\!=\!4096$), and 
GritLM-Mistral-unified-7B ($d\!=\!4096$) ~\cite{muennighoff2024generative}. 
For brevity, we refer to Qwen-Embedding models as Qwen and LLM2Vec models as L2V throughout the remainder of the paper.

We L2-normalize query and document embeddings once. After applying the query mask, we do not re-normalize the query; this only rescales scores by a query-aware constant and does not affect ranking.

\paragraph{Datasets and Splits}
Experiments are conducted on SciFact, NFCorpus, and MS MARCO. We use BEIR-provided in-domain splits (train/test) for each dataset, without cross-domain transfer. For MS MARCO, we subsample 50{,}000 query–document pairs for training, and for all datasets we reserve 10\% of the training set as a validation set.

\paragraph{Training and Inference}
Our predictor is a single fully connected layer mapping $\mathbb{R}^D \!\to\! \mathbb{R}^D$, followed by a (log-)softmax to produce a query-aware importance distribution $\hat{\pi}_q$ over dimensions. 
Training targets $\pi_q$ are constructed from labeled relevance and hard negatives: we first rank non-relevant documents by similarity $s(d)=e_d\!\cdot\! e_q$, take the top-$K$ highest-scoring items, and uniformly sample $M$ of them to form the hard-negative set $D^{-}(q)$; $K$ and $M$ are tunable hyperparameters. 
The model is optimized with AdamW and cosine-annealing learning rate using the KL divergence objective $\mathrm{KL}(\pi_q \,\|\, \hat{\pi}_q)$, and we select the best checkpoint by validation KL.

At inference time, the underlying encoder and similarity function remain unchanged; our method only selects dimensions. 
In our experiments, we evaluate $k$ over a grid from 2\% to 100\% of $D$ in steps of 2\%, and report the peak NDCG@10 achieved for each method and model.

\paragraph{Baselines}
We compare against the following baselines.
\textbf{Full}: a no-truncation baseline that uses the full $D$-dimensional embeddings.
\textbf{Cutoff}: a static prefix that keeps the first-$k$ coordinates of the original embedding.
\textbf{Norm}: a query-aware baseline that scores dimensions by the absolute value of the query coordinate and retains the top-$k$ dimensions after sorting by $|e_{q,j}|$.
\textbf{Adapter}~\cite{yoon2024search}: our implementation of the supervised search adapter , 
which learns a projection on top of the frozen encoder and is reported only at full 
$D$ dimensions, as it is trained to optimize full-dimensional retrieval performance.
\textbf{DIME-PRF}~\cite{faggioli2024dimension}: our reproduction of DIME, which uses pseudo-relevance feedback with the top-1 retrieved document as a pseudo-positive to estimate per-query dimension importance, followed by top-$k$ selection.
\textbf{Eclipse-PRF}~\cite{d2025eclipse}: our reproduction of Eclipse, which extends DIME-PRF by additionally incorporating pseudo-negatives from the PRF set for contrastive dimension scoring.
For both DIME-PRF and Eclipse-PRF, we explored different pseudo-relevance feedback configurations and found that using the top-1 document as pseudo-positive yields the best performance.

We do not include methods that explicitly optimize performance under reduced dimensionality (e.g., PCA or multi-level MRL adapters \cite{yoon2024matryoshka,zhang2025smec}) as baselines, since our primary goal is to improve the peak effectiveness at the full embedding dimensionality. Such dimensionality-reduction approaches are designed to preserve performance as dimensions are truncated, but their reduced representations are not expected to surpass the effectiveness of the original full-dimensional embeddings.

\paragraph{Hyperparameters and Tuning}
We tune per model–dataset on the validation split by selecting the configuration that maximizes the sum of NDCG@10 across retained embedding dimensions $\{D, D/2, D/4, D/8\}$. The temperature $\tau$ for the importance softmax is tuned via a grid search, while the negative sampling parameters are fixed to $K{=}1000$ (hard-negative pool size) and $M{=}64$ (number of in-batch negatives).

\section{Experiments}
\subsection{Main Results}
Table~\ref{tab:main_all_models} summarizes peak retrieval effectiveness (NDCG@10) and the corresponding fraction of retained dimensions (in parentheses). Each cell shows the best score achieved over all tested values of $k$, together with the fraction $k / D$ in parentheses.

Across datasets and base encoders, our query-aware predictor consistently matches or improves upon the full-dimensional baseline while using substantially fewer dimensions, and it often surpasses both supervised adapter tuning (\textbf{Adapter}) and PRF-based query-aware masking (\textbf{DIME}/\textbf{Eclipse}).

In contrast, heuristic truncation strategies like keeping a fixed prefix (\textbf{Cutoff}) or selecting dimensions by query norm (\textbf{Norm}) yield limited or inconsistent gains over the baseline, suggesting that the improvements are driven by learned, query-aware dimension scoring rather than truncation alone or simple magnitude-based selection.

Notably, our best results typically arise at roughly 20–40\% of retained dimensions, suggesting that many embedding coordinates are not merely redundant but unhelpful for retrieval; removing them via query-aware masking can improve effectiveness over using all dimensions.

Finally, using a single fixed retention rate of 30\% (\textbf{Ours}@30\%) achieves performance close to the per-setting peak across models and datasets, which we adopt as a practical default for deployment-oriented evaluation.
\subsection{Effect of Retained Dimension Ratio}
\begin{figure}[tbp]
  \centering
  \includegraphics[width=\linewidth]{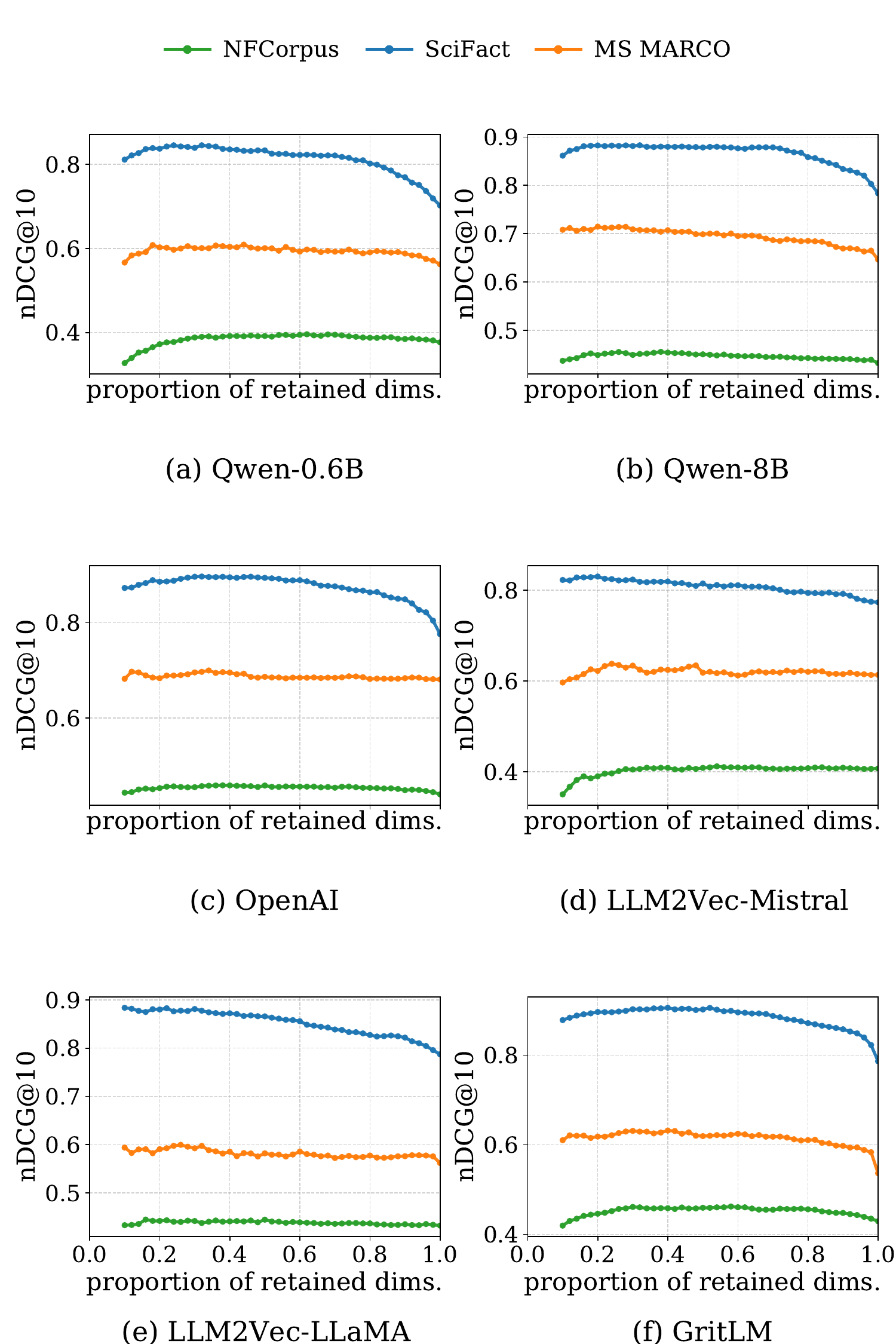}
  \caption{
NDCG@10 as a function of retained dimension ratio $k/D$. Each panel corresponds to one model; our method consistently reaches a peak and plateau around 20--50\% of the dimensions.
  }
    \label{fig:retention-curves}
\end{figure}

Figure~\ref{fig:retention-curves} visualizes retrieval effectiveness as a function of the retained dimension ratio $k / D$ for the six  representative models. 

Across all models, our approach exhibits a consistent pattern: effectiveness reaches a pronounced peak and plateau when retaining roughly 20–40\% of the dimensions. 
Within this range, our method typically matches or exceeds the full-dimensional baseline while using substantially fewer coordinates. 
Beyond 50\%, additional dimensions bring little or no benefit, and in some cases slightly hurt performance, suggesting that many dimensions in standard dense embeddings are redundant or even detrimental for retrieval. 
This broad plateau indicates that the predictor concentrates query information on a compact, high-value subset of dimensions while safely discarding less informative components, and implies that in practice one can obtain near-peak effectiveness with any retained dimensions in the 20–40\% range, even without knowing the exact optimal $k$.
\section{Discussion}
\subsection{Ablation of Oracle Dimension Importance}

\begin{figure}[tbp]
  \centering
  \includegraphics[width=\linewidth]{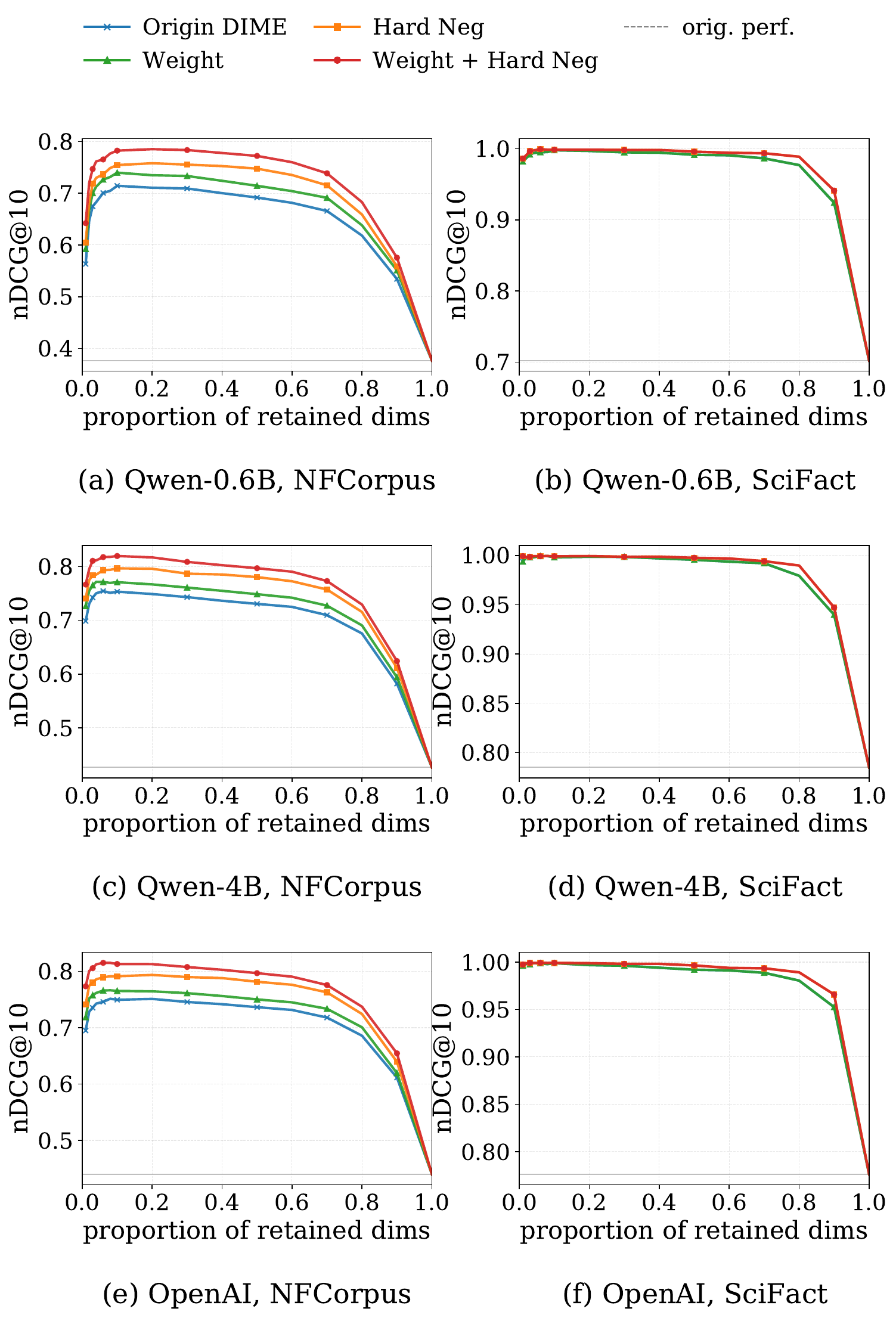}
  \caption{
Oracle-style dimension selection improves upper bounds, with gains peaking near 20\% and further boosted by positive weighting and hard negative mining.
  }
  \label{fig:ablation_oracle}
\end{figure}

DIME has shown that, given labels, selecting a subset of dimensions per query can significantly improve retrieval performance. Because the original study used different models and datasets, we perform a minimal re-validation: on SciFact (single-level labels) and NFCorpus (multi-level labels), we re-test the key trends using Qwen and OpenAI models, respectively. Building on DIME, we further introduce positive example weighting and hard negative mining, and present the ablation results in Figure~\ref{fig:ablation_oracle}. 

Results indicate that the oracle gain peaks when selecting roughly 20\% of dimensions and remains stable when selecting 20\%--80\% of dimensions, consistent with the trend reported in the DIME paper. Moreover, our positive weighting and hard negative mining further improve this upper bound.

\subsection{Robustness to Hyperparameters}

We assess robustness to both stochasticity and hyperparameter choices across all model--dataset pairs, and provide representative sensitivity curves for two models (Qwen-8B and GritLM) and two datasets (SciFact and NFCorpus); other combinations exhibit consistent trends. Full results are reported in Appendix~\ref{app:sen}. 

\paragraph{Random-Seed Sensitivity.}
We fix all hyperparameters to their default settings and run 10 independent trainings with distinct seeds for each model--dataset pair. The observed variation in NDCG@10 is minor, indicating that our method is largely insensitive to initialization randomness and other stochastic factors.

\paragraph{Hyperparameter Sensitivity.}
We conduct controlled one-at-a-time sweeps on Qwen-8B with SciFact, varying one hyperparameter at a time while fixing the others to their defaults. Specifically, we sweep the number of training epochs, the temperature $\tau$, the hard-negative pool size $K$, and the number of in-batch negatives $M$, evaluating each setting across a fixed set of retained dimensionalities (Appendix~\ref{app:sen}). We find that $K$ and $M$ have only modest impact across dimensionalities, so we adopt conservative defaults of $K{=}1000$ and $M{=}64$. The best performance is obtained around $\tau{=}0.01$, and performance typically saturates after roughly 100 epochs.

A complete list of hyperparameters and search ranges is provided in Appendix~\ref{tab:hyperparam_ranges}.

\subsection{Dimension-Selection Consistency Analysis}
We study whether the dimension importance predictor is consistent—similar queries select similar dimensions and dissimilar queries select markedly different ones—thereby supporting interpretability. Concretely, for each model (Qwen-4B, Qwen-8B, OpenAI, GritLM) on SciFact and NFCorpus, we consider all unordered query pairs within the test split at a fixed retained dimensionality $k=512$. For a query $q$ with frozen embedding $e_q \in \mathbb{R}^D$ and predictor output $\hat{\pi}_q$, we define the selected index set $T_k(q)=\mathrm{Top}\text{-}k(\hat{\pi}_q)$ and compute (i) query similarity $s_k(q,q')$ and (ii) Jaccard similarity of selected dimensions
$J_k(q,q')=\frac{|T_k(q)\cap T_k(q')|}{|T_k(q)\cup T_k(q')|}$. We then report the Pearson correlation between $\{s_k(q,q')\}$ and $\{J_k(q,q')\}$ over all pairs in each test set.

\begin{table}[t]
  \centering
  \caption{Pearson correlation between query similarity $s_k$ and Jaccard similarity $J_k$ of selected dimensions ($k=512$) on \emph{SciFact} and \emph{NFCorpus} test sets.}
  \label{tab:pearson_dim_consistency}
  \begin{tabular}{lcc}
    \toprule
    \textbf{Model} & \textbf{Dataset}  & \textbf{Pearson} \\
    \midrule
    Qwen-4B  & SciFact   & 0.541 \\
    Qwen-8B  & SciFact   & 0.415 \\
    OpenAI   & SciFact   & 0.498 \\
    GritLM   & SciFact  & 0.478 \\
    Qwen-4B  & NFCorpus  & 0.335 \\
    Qwen-8B  & NFCorpus  & 0.361 \\
    OpenAI   & NFCorpus  & 0.387 \\
    GritLM   & NFCorpus  & 0.319 \\
    \bottomrule
  \end{tabular}
\end{table}

Across both domains, we observe moderate positive correlations (typically in the 0.3–0.5 range), indicating that similar queries tend to share important dimensions and that the predictor captures this structure on unseen test queries.
\subsection{Train with LLM-Generated Queries}
\begin{table}[t]
\centering
\small
\setlength{\tabcolsep}{5pt}
\caption{
Comparison of the best performance achievable with 
\textbf{Full} embeddings,
\textbf{Unsup-LLM} predictors trained with LLM-generated pseudo-queries,
and \textbf{Supervised} predictors trained with relevance labels.
Numbers are NDCG@10 on the test split; 
best result per model/dataset is in bold.
}
\begin{tabular}{l l c c c}
\toprule
\textbf{Model} & \textbf{Dataset} & \textbf{Full} & \textbf{Unsup} & \textbf{Supervised} \\
\midrule
\multirow{3}{*}{GritLM} 
 & SciFact   & 0.786 & 0.790 & \textbf{0.902} \\
 & NFCorpus  & 0.439 & 0.441 & \textbf{0.459} \\
 & MS MARCO  & 0.536 & 0.602 & \textbf{0.626} \\
\midrule
\multirow{3}{*}{Qwen-8B} 
 & SciFact   & 0.783 & 0.785 & \textbf{0.899} \\
 & NFCorpus  & 0.432 & 0.435 & \textbf{0.455} \\
 & MS MARCO  & 0.646 & 0.697 & \textbf{0.715} \\
\midrule
\multirow{3}{*}{OpenAI} 
 & SciFact   & 0.776 & 0.778 & \textbf{0.897} \\
 & NFCorpus  & 0.440 & 0.441 & \textbf{0.459} \\
 & MS MARCO  & 0.681 & 0.694 & \textbf{0.700} \\
\bottomrule
\end{tabular}

\label{tab:unsup-llm-analysis}
\end{table}

To assess whether dimension importance predictors can be trained without human relevance labels, we construct a synthetic training set by generating pseudo-queries from documents using an LLM (\textit{doc}~$\rightarrow$~\textit{LLM}~$\rightarrow$~\textit{query}). Each pseudo-query is paired with its source document as a positive example, and negatives are sampled from the remaining corpus. We then train predictors on these LLM-generated pairs using the same oracle construction and loss as in the supervised setting.

For evaluation, we follow the protocol used in our main experiments: we perform a 2\%-step sweep over the retained dimension ratio and, for each model/dataset, report the best NDCG@10. The resulting upper-bound performance for the three predictors is summarized in Table~\ref{tab:unsup-llm-analysis}.

Empirically, on the large, web-style MS MARCO benchmark, predictors trained from LLM-generated pseudo-queries achieve non-trivial gains over the full-dimensional baseline and recover a substantial portion of the improvements obtained by fully supervised predictors. In contrast, on smaller, domain-specific collections such as SciFact and NFCorpus, the same unsupervised training yields only marginal improvements. This suggests that synthetic queries are most effective when the target query distribution is close to generic natural-language questions and ample data is available, whereas high-quality task-specific labels remain crucial for learning fine-grained dimension importance in specialized low-resource settings.

\subsection{Adapters with Query-Aware Selection}

Given that adapters and PRF-based methods are complementary in spirit—adapters improve the global quality of the embedding space, while methods such as DIME-PRF, Eclipse-PRF, and ours perform query-aware dimension selection—it is natural to ask whether stacking them could further improve performance. Concretely, we experimented with two hybrid configurations: (i) applying DIME-PRF or Eclipse-PRF on top of a supervised search adapter, and (ii) training our dimension selector on adapter outputs instead of on the base encoder. 

\begin{table}[t]
\centering
\small
\caption{
Effect of combining adapters with query-aware dimension selection.
Numbers are Peak NDCG@10 on two representative datasets and models.
A = Adapter, E = Eclipse, O = Ours; thus A+E and A+O denote their combinations. }
\label{tab:adapter_prf_composition}
\begin{adjustbox}{max width=0.48\textwidth}
\begin{tabular}{lcccc}
\toprule
& \multicolumn{2}{c}{MS MARCO} & \multicolumn{2}{c}{SciFact} \\
\cmidrule(lr){2-3} \cmidrule(lr){4-5}
Method & Qwen-8B & GritLM & Qwen-8B & GritLM \\
\midrule
Baseline                 & 0.646 & 0.536 & 0.783 & 0.786 \\
A                  & 0.698 & 0.593 & 0.883 & 0.883 \\
E                  & 0.668 & 0.565 & 0.785 & 0.787 \\
O                     & 0.714 & 0.632 & 0.883 & \textbf{0.906} \\
A + E                    & 0.704 & 0.605 & \textbf{0.883} & 0.883 \\
A + O                    & \textbf{0.731} & \textbf{0.637} & 0.883 & 0.883 \\
\bottomrule
\end{tabular}
\end{adjustbox}
\end{table}

Table~\ref{tab:adapter_prf_composition} reports results on two representative datasets and models. On MS MARCO, both Eclipse and our method provide additional gains when applied on top of the adapter, with \textit{Adapter+Ours} achieving the best overall effectiveness. However, on SciFact our selector can slightly outperform the adapter, whereas \textit{Adapter+Ours} reverts to the adapter’s effectiveness. 

\begin{table}[t]
\centering
\caption{Average Jaccard similarity of selected important dimensions between query pairs, computed over 20{,}000 sampled query pairs per dataset. Higher values indicate greater overlap. A = Adapter, O = Ours; A+O denote their combinations.}
\begin{adjustbox}{max width=\linewidth}
\begin{tabular}{lcccc}
\toprule
& \multicolumn{2}{c}{MS MARCO} & \multicolumn{2}{c}{SciFact} \\
\cmidrule(lr){2-3} \cmidrule(lr){4-5}
Method & Qwen-8B & GritLM & Qwen-8B & GritLM \\
\midrule
O   & 0.369 & 0.360 & 0.339 & 0.336 \\      
A + O  & 0.392 & 0.397 & 0.479 & 0.506 \\      
\bottomrule
\end{tabular}
\end{adjustbox}

\label{tab:query_jaccard}
\end{table}

To better understand this behavior, we analyze how adapters affect the distribution of important query dimensions. Specifically, for GritLM and Qwen-8B on MS~MARCO and SciFact, we measure the Jaccard similarity between the sets of top-$2048$ dimensions selected by our selector. For each dataset, we sample 20{,}000 query pairs $(q_i, q_j)$ from the training set and compute the Jaccard similarity between their important-dimension sets, both with and without the adapter. 

As shown in Table~\ref{tab:query_jaccard}, without adapters the average Jaccard similarity is comparable across datasets. After adding adapters, MS~MARCO exhibits only a modest change, whereas SciFact shows a substantial increase in overlap. This indicates that, on SciFact, the adapter makes the router’s notion of “important dimensions” more homogeneous across queries, reducing the need for additional query-wise specialization. In contrast, MS~MARCO remains more heterogeneous, which explains why our method can still bring gains on top of the adapter.

\section{Conclusion}
We introduced a query-aware framework for learning which embedding dimensions matter most for relevance. Our method distills dimension importance from supervised labels into a compact predictor that produces dynamic per-query masks at inference, enabling similarity computations to focus on the most informative coordinates while preserving or improving ranking quality. This approach captures query-aware importance, better aligning the retained representation with the signals that drive relevance for each information need.

\section{Limitation}
Our approach has two main limitations. First, it relies on supervised relevance signals to construct oracle dimension-importance scores; even with LLM-generated pseudo-queries, high-quality labels remain important, especially on small, domain-specific datasets. Second, our method only operates at the level of per-query dimension selection over frozen embeddings and cannot improve the underlying encoder itself, so its overall effectiveness is inherently bounded by the quality of the base dense retriever.

\bibliography{ref}
\appendix
\section{Appendix}
\label{sec:appendix}

\subsection{Hyperparameters and Randomness Sensitivity}

\label{app:sen}

For hyperparameter sensitivity, we perform controlled one-at-a-time sweeps on \textbf{Qwen-8B} with \textbf{SciFact}, varying one hyperparameter while fixing the others to their defaults. Concretely, we vary: \textbf{epochs} = \{20, 30, 50, 100, 200\} (default 100), temperature $\tau$ = \{0.005, 0.01, 0.02, 0.05, 0.1\} (default 0.01), hard-negative pool size $K$ = \{500, 1000, 2000, 3000\} (default 1000), and number of in-batch negatives $M$ = \{16, 32, 64, 128, 256\} (default 64). For each setting, we evaluate performance across a fixed set of embedding dimensionalities.

\begin{figure}[htbp]
  \centering
  \includegraphics[width=\linewidth]{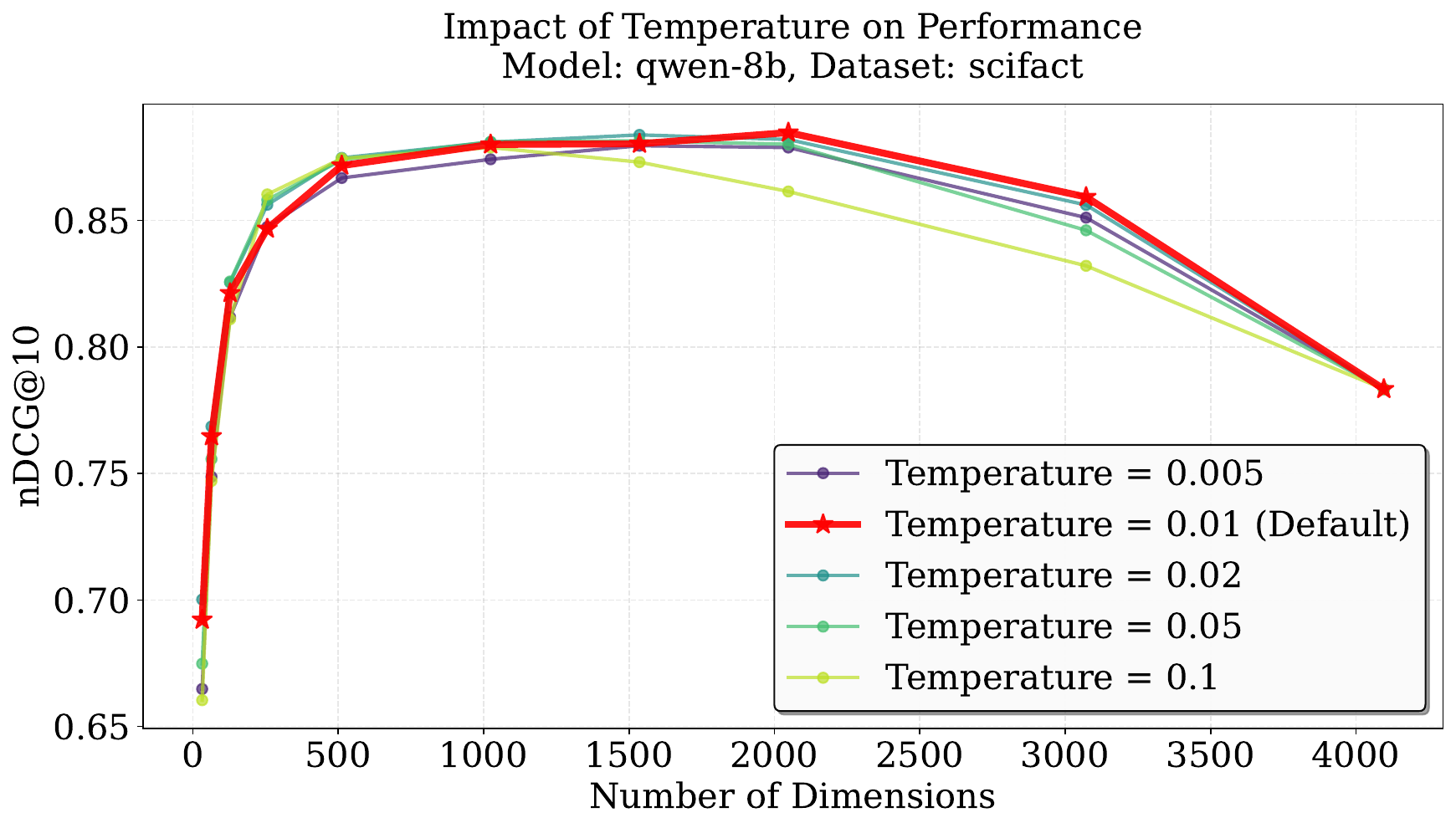}
  \caption{
Effect of temperature.
  }
\end{figure}
\begin{figure}[htbp]
  \centering
  \includegraphics[width=\linewidth]{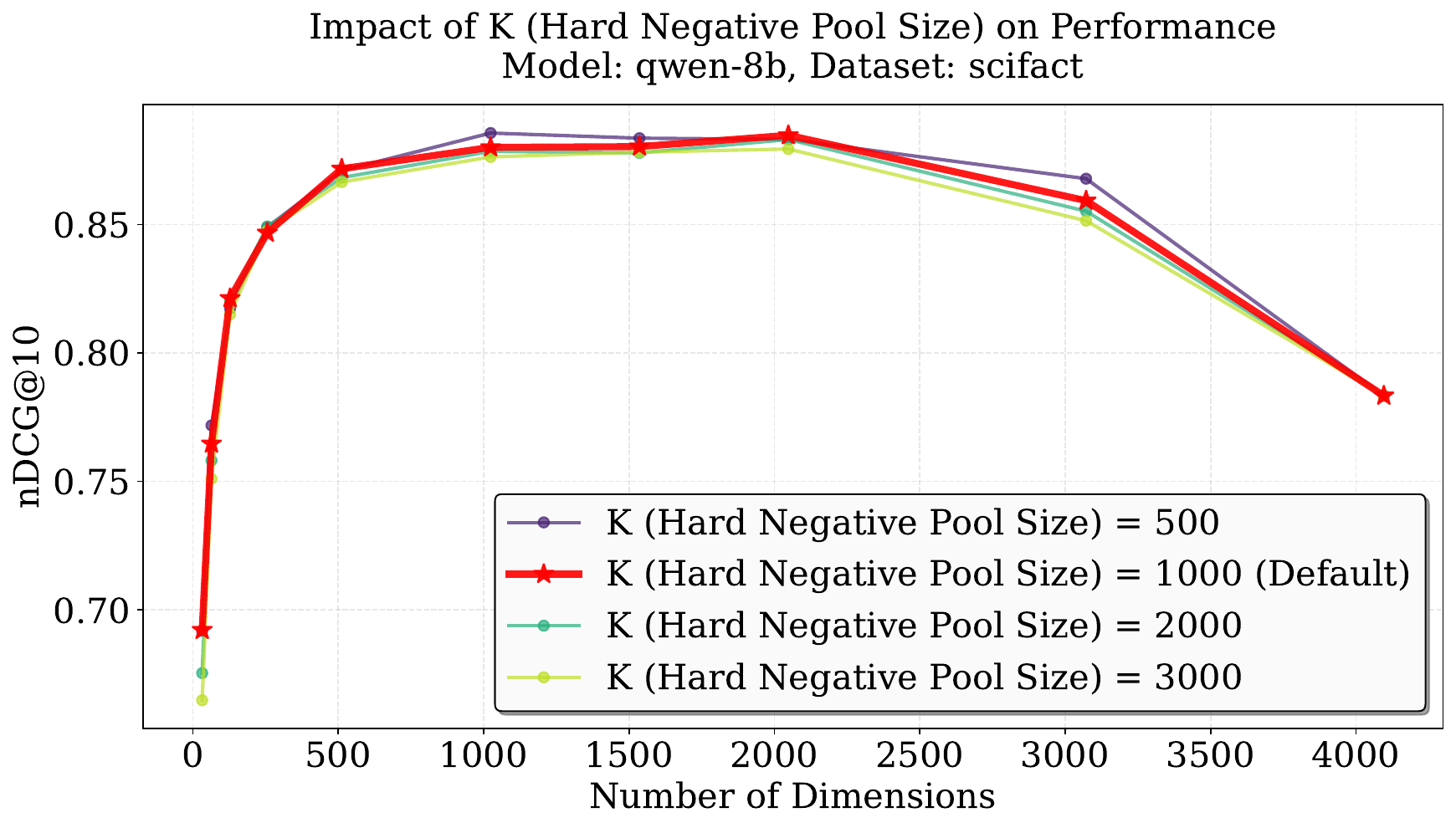}
  \caption{
Effect of K.
  }
\end{figure}
\begin{figure}[htbp]
  \centering
  \includegraphics[width=\linewidth]{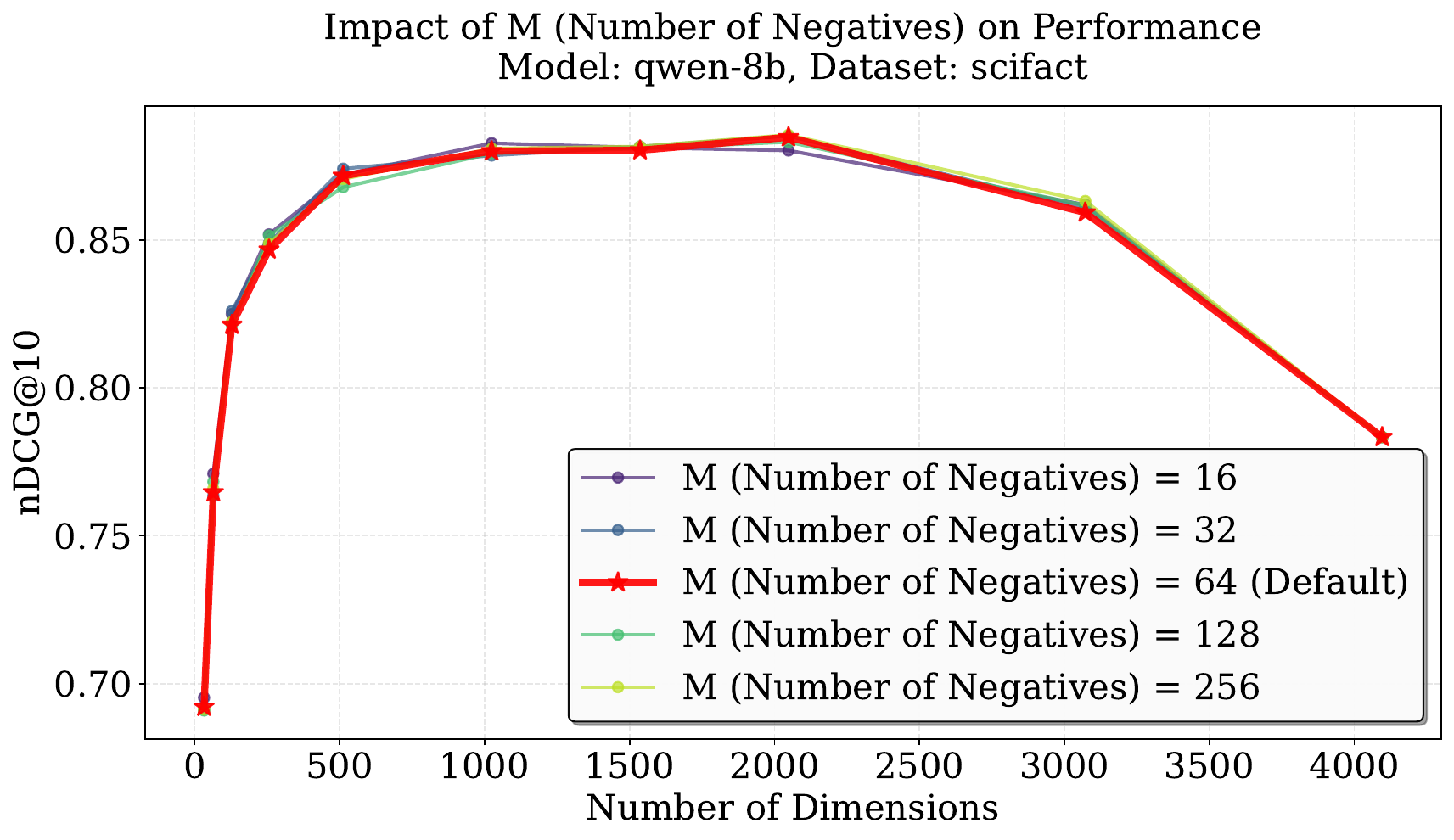}
  \caption{
Effect of M.
  }
\end{figure}
\begin{figure}[htbp]
  \centering
  \includegraphics[width=\linewidth]{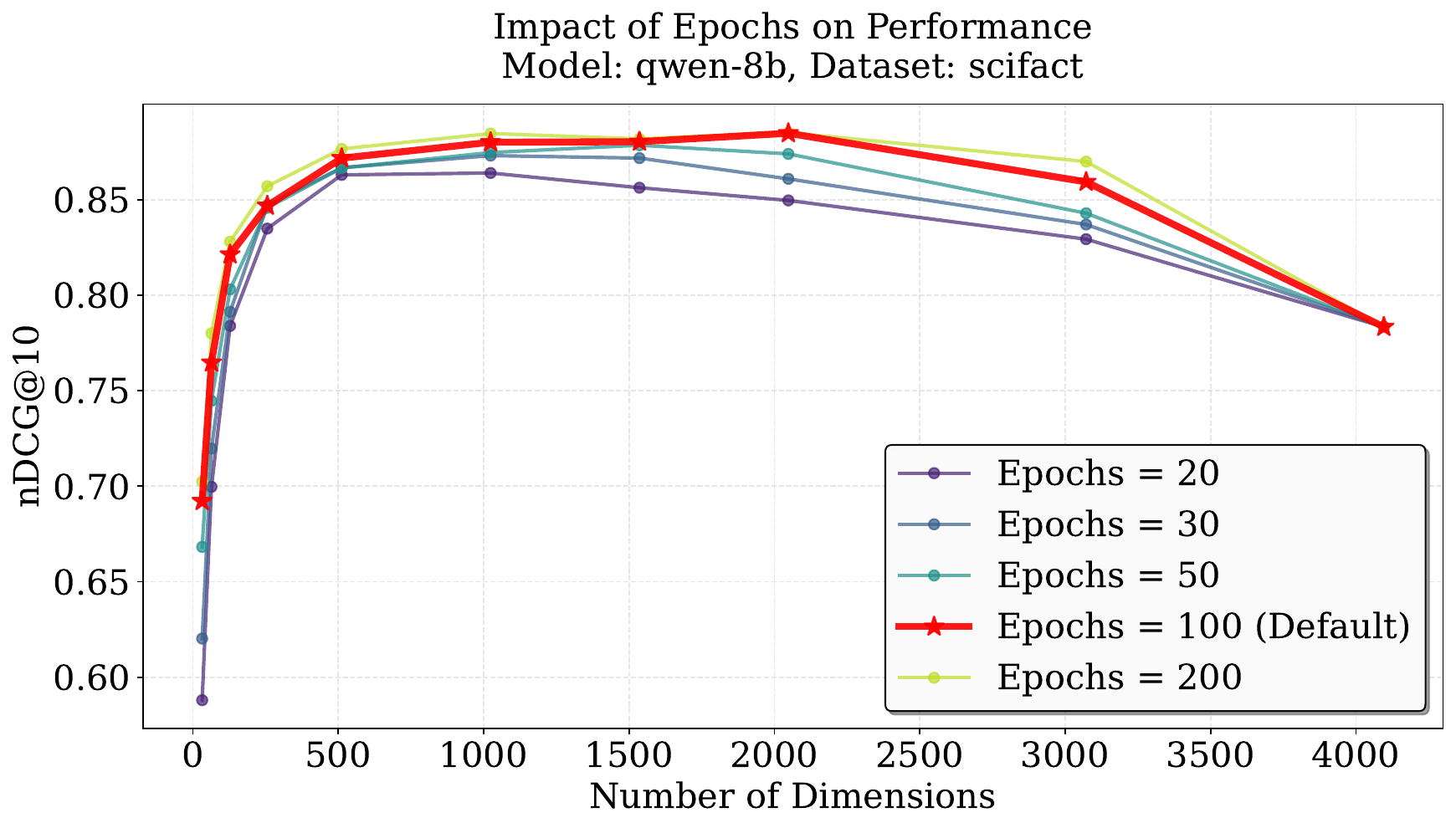}
  \caption{
Effect of epoch.
  }
\end{figure}

For random-seed sensitivity, we fix all hyperparameters to their default settings and run 10 independent trainings with distinct seeds for each model–dataset pair. The observed variation in NDCG@10 is marginal, indicating that our method is largely insensitive to initialization randomness and other stochastic factors. 

\begin{figure}[htbp]
\centering
\includegraphics[width=\linewidth]{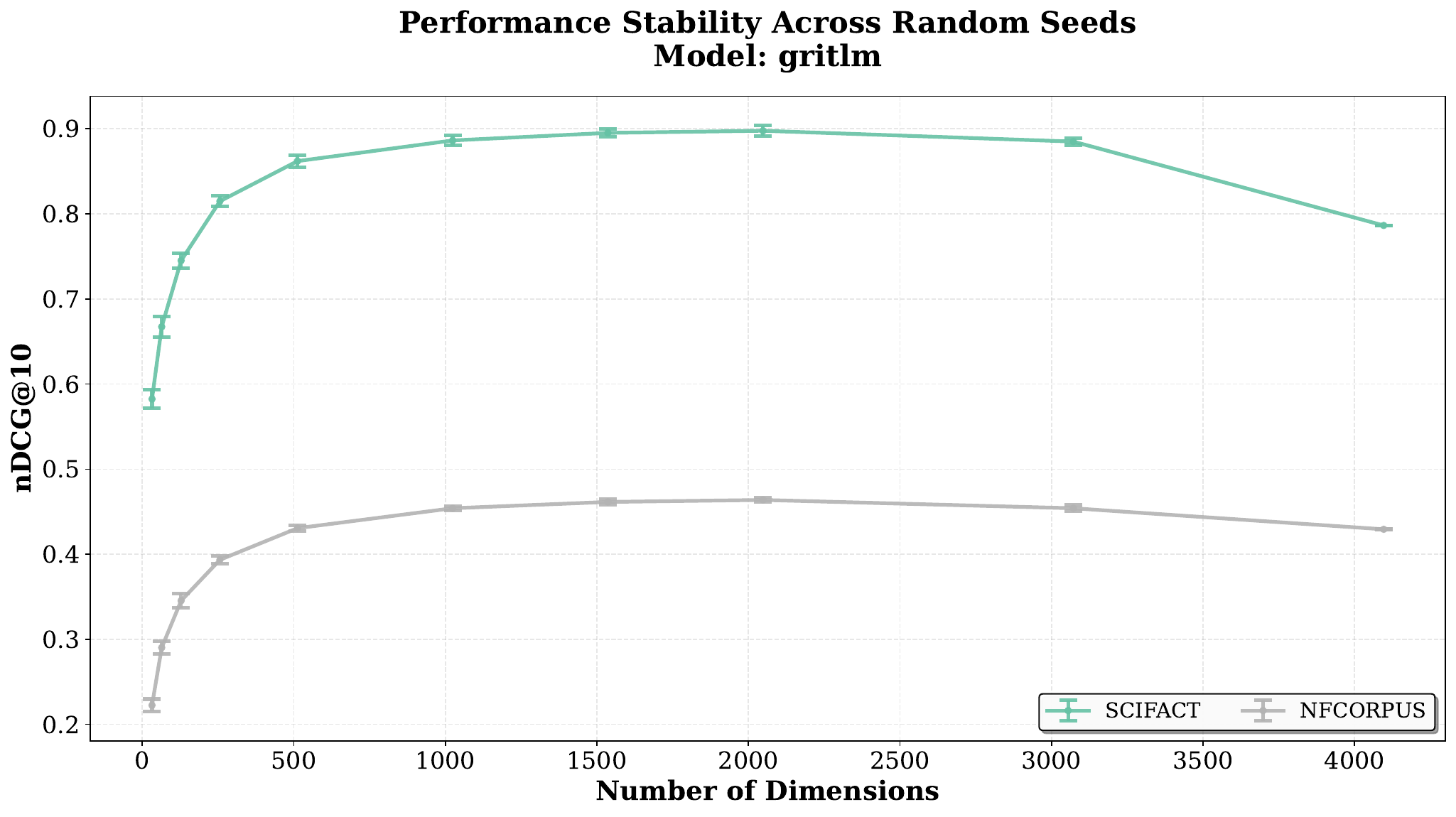}
\vspace{0.6em}
\includegraphics[width=\linewidth]{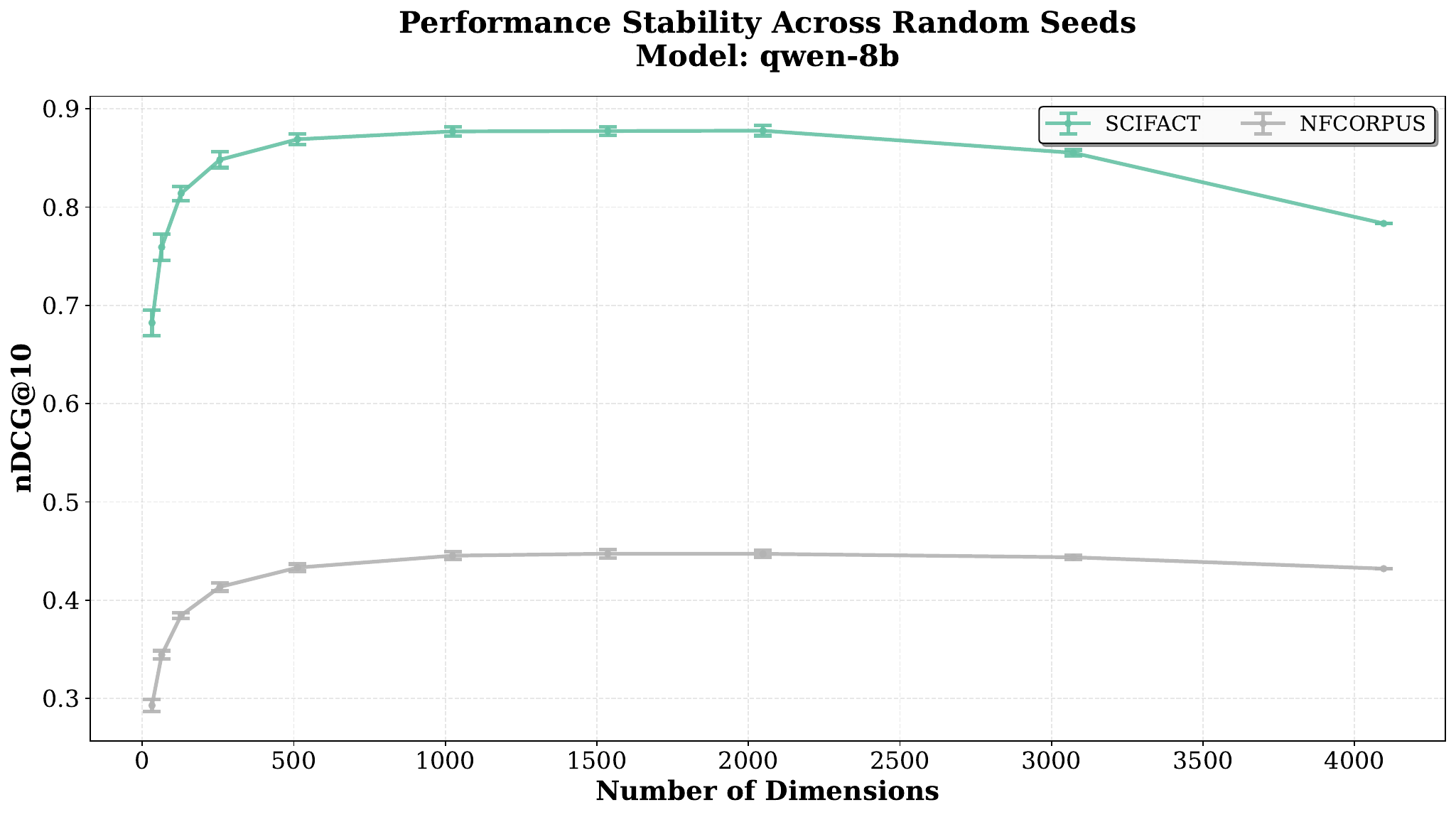}
\caption{Random-seed sensitivity on SciFact (top) and NFCorpus (bottom) with qwen-8b and gritlm. Error bars denote $\pm 1$ standard deviation of NDCG@10 across seeds, showing marginal variability and overall robustness.}
\label{fig:seed_sensitivity}
\end{figure}

\subsection{Hyperparameters Details}

\begin{table}[htbp]
\centering
\small
\setlength{\tabcolsep}{6pt}
\renewcommand{\arraystretch}{1.15}
\caption{Hyperparameter search ranges used in our experiments.}
\label{tab:hyperparam_ranges}
\begin{adjustbox}{max width=\textwidth}
\begin{tabular}{@{}l l@{}}
\toprule
\textbf{Hyperparameter} & \textbf{Search range / Value} \\
\midrule
\multicolumn{2}{@{}l}{\textit{Searched (model selection on validation set)}} \\
\midrule
Temperature $\tau$ &
\makecell[l]{\{0.0001, 0.0002, 0.0005, 0.001, 0.002,\\
0.005, 0.01, 0.02, 0.05\}} \\
Training epochs &
\{20, 30, 50, 75, 100, 200\} \\
\midrule
\multicolumn{2}{@{}l}{\textit{Fixed (shared across all models and datasets)}} \\
\midrule
Optimizer & AdamW \\
Learning rate & 1e{-}4 \\
Weight decay & 0.01 \\
Batch size & 256 \\
Dropout & 0.1 \\
Hard-negative pool size $K$ & 1000 \\
In-batch negatives $M$ & 64 \\
\bottomrule
\end{tabular}
\end{adjustbox}
\end{table}

\end{document}